\documentstyle[12pt]{article}
\newcommand{\dmsq}{$\Delta m^2$}
\newcommand{\mixing}{$\sin^2 2\theta$}

\newcommand {\omutau} {$\nu_\mu \rightarrow \nu_\tau$}
\newcommand {\omuste} {$\nu_\mu \rightarrow \nu_{\mathrm{s}}$}
\newcommand {\omue} {$\nu_\mu \rightarrow \nu_e$}

\newcommand {\xmudec} {$\tau^- \rightarrow \mu^-\nu\bar{\nu}$}

\def\xa1dec{$\tau^- \rightarrow a_1^-\nu$}
\def\za1dec{$\tau^- \rightarrow \pi^- \pi^+ \pi^- \nu$}
\def\za1op{$\tau^- \rightarrow \pi^- \pi^0 \pi^0 \nu$}
\def\yrho_2pi{$\rho^- \rightarrow \pi^-\pi^0$}
\def\ya1_3pi{$a_1^- \rightarrow \pi^-\pi^+\pi^-$}
\def\ypi_gg{$\pi^0 \rightarrow \gamma\gamma$}

\begin {document}
\title
{Detecting $\nu_\tau$ appearance in the spectra of quasielastic CC events}
\author
{A.E. Asratyan\thanks{Corresponding author. Tel.: 095-237-0079. E-mail
address: asratyan@vxitep.itep.ru.},
G.V. Davidenko, A.G. Dolgolenko, V.S. Kaftanov,\\
M.A. Kubantsev\thanks{Now at Fermi National Accelerator Laboratory, Batavia, 
IL 60510, USA.},
and V.S. Verebryusov\\
\normalsize{\it Institute of Theoretical and Experimental Physics,}
\normalsize{\it Moscow 117259, Russia}}
\date {\today}
\maketitle

\begin{abstract}
A method for detecting the transition \omutau\ in long-baseline accelerator
experiments, that consists in comparing the far-to-near ratios of the spectra
of quasielastic CC events generated by  high- and low-energy beams of muon 
neutrinos, is proposed. The test may be accessible to big calorimeters with 
muon spectrometry like MINOS, and is limited by statistics rather than 
systematics.
\end{abstract}
PACS: 14.60.Pq; 14.60.Fg
\\Keywords: Neutrino oscillations, $\nu_\tau$ appearance
\clearpage

     The data of Super-Kamiokande \cite{sk_on_omutau} favor the transition
\omutau\ as the source of the deficit of muon neutrinos from the atmosphere.
However, this still has to be verified by directly observing $\nu_\tau$
appearance in accelerator long-baseline experiments. The options discussed
thus far, all involving fine instrumentation on a large scale, are to detect 
the secondary $\tau$ by range in emulsion \cite{opera}, by Cherenkov light 
\cite{cherenkov}, or by the transverse momentum carried away by the decay 
neutrino(s) \cite{icarus}. By contrast, in this paper we wish to formulate
a $\tau$ signature that is solely based on the energy spectra of CC events,
and therefore should be accessible to relatively coarse calorimeters with
muon spectrometry. We assume that the experiment includes a near detector of 
the same structure as the far detector, irradiated by the same neutrino beam 
but over a short baseline that rules out any significant effects of neutrino
oscillations \cite{minos}. Thereby, the systematic uncertainties in comparing
the interactions of primary and oscillated neutrinos are largely eliminated.
Our aim is to distinguish the muonic decays of $\tau$ leptons against the 
background of $\nu_\mu$-induced CC events. In order to minimize the effects 
of $\nu_\mu$ disappearance, the data collected with a harder $\tau$-producing
beam are compared with those for a softer reference beam in which $\tau$ 
production is suppressed by the threshold effect. The analysis is restricted 
to quasielastics (QE), that is, to neutrino events featuring a muon and small 
hadronic energy.

     As soon as the first maximum of the oscillation lies below the mean
energy of muon neutrinos in the beam, or
$\Delta m^2 L / \langle E_\nu \rangle < 1.24$ eV$^2$km/GeV,
much of the signal from QE production and muonic decay of the $\tau$ is
at relatively low values of visible energy $E$. That is because the tau 
neutrinos arising from \omutau\ are softer on average than muon neutrinos, 
the threshold effect is relatively mild for quasielastics, and a large 
fraction of incident energy is taken away by the two neutrinos from \xmudec. 
Let $f(E)$ be the spectrum of QE events observed in the far detector,
$n(E)$---the spectrum of similar events in the near detector that has been 
extrapolated and normalized to the location of the far detector, and 
$R(E)$---the ratio of the two: $R(E) = f(E) / n(E)$. In the case of $\nu_\mu$
disappearance through the transitions \omue\ or \omuste, where 
$\nu_{\mathrm{s}}$ is the hypothesized sterile neutrino, the ratio $R$ for 
the harder beam should be identically equal to that for the softer beam: 
$R_{\mathrm{hard}}(E) = R_{\mathrm{soft}}(E)$. However, in the case of 
\omutau\ this equation is violated by the process of $\tau$ production and 
muonic decay, that predominantly occurs in the harder beam and shows up as a 
low-$E$ enhancement of the corresponding "far" spectrum 
$f_{\mathrm{hard}}(E)$. This causes the ratio $R_{\mathrm{hard}}$ to exceed 
$R_{\mathrm{soft}}$ towards low values of visible energy $E$. The latter 
effect, that may provide a specific signature of $\nu_\tau$ appearance, is 
investigated in this paper.

     The simulation assumes that the MINOS detector \cite{minos}, a 5.4-kton
iron--scintillator calorimeter under construction in the Soudan mine in
Minnesota, is irradiated by neutrinos from Fermilab over a baseline of 
730 km. In this detector, secondary muons will be sign-selected by curvature
in magnetic field and momentum-analyzed by range and/or curvature. In the
NuMI program due to start in 2003 \cite{numi}, the experiment will use a
variety of neutrino beams generated by the Main Injector (MI), a 120-GeV
proton machine. At a later stage, MINOS may be irradiated by a Neutrino
Factory (NF) at Fermilab \cite{nufactory}. For the soft reference beam of 
muon neutrinos, we always select the MI low-energy beam foreseen by the NuMI
program \cite{numi}. The harder $\tau$-producing $\nu_\mu$ beam is assigned 
either as the MI high-energy beam \cite{numi},
or as that generated by a neutrino factory storing 
20-GeV negative muons. The mean $\nu_\mu$ energies in the MI low-energy, MI 
high-energy, and NF beams are 5, 12, and 14 GeV, respectively. We assume that
the neutrino factory delivers 10$^{20}$ useful $\mu^-$ decays per year of 
operation. Apart from higher intensity, an important advantage of the NF beam
compared to the MI high-energy beam is that less of the total $\nu_\mu$ flux 
is at low energies.

     Charged-current interactions of the $\nu_\mu$ and $\nu_\tau$ are 
generated using the NEUGEN package that is based on the Soudan-2 Monte 
Carlo \cite{generator}. The response of the detector is not 
simulated in full detail; instead, the resolution in muon energy is 
approximated as $\delta E_\mu = 0.11 \times E_\mu^{\mathrm{true}}$ and in 
energy transfer to hadrons---as $\delta\nu$ (GeV) = 
$0.55 \times \sqrt{\nu^{\mathrm{true}}}$ \cite{resolutions}. Quasielastic 
events are selected as those with $E_\mu > 800$ MeV and $\nu < 1$ GeV, where 
$E_\mu$ and $\nu$ are the smeared values of muon energy and of energy 
transfer to hadrons, respectively\footnote{These selections should be 
viewed as illustrative. The actual selections will be based on a detailed
simulation of detector response to CC events with small hadronic energy.}.
Given the characteristic topology of such events in the detector (a single 
track traversing more than three nuclear interaction lengths in iron plus a
few scintillator hits near the primary vertex), we assume that they are 
reconstructed with 100\% efficiency and that the background from pion
punchthrough is insignificant. The visible energy $E$ of a CC event is again
estimated in terms of smeared quantities: $E = E_\mu + \nu$. Systematic 
uncertainties of the near spectra $n(E)$ are neglected.

\begin{figure}[p]
\vspace{18 cm}
\includegraphics{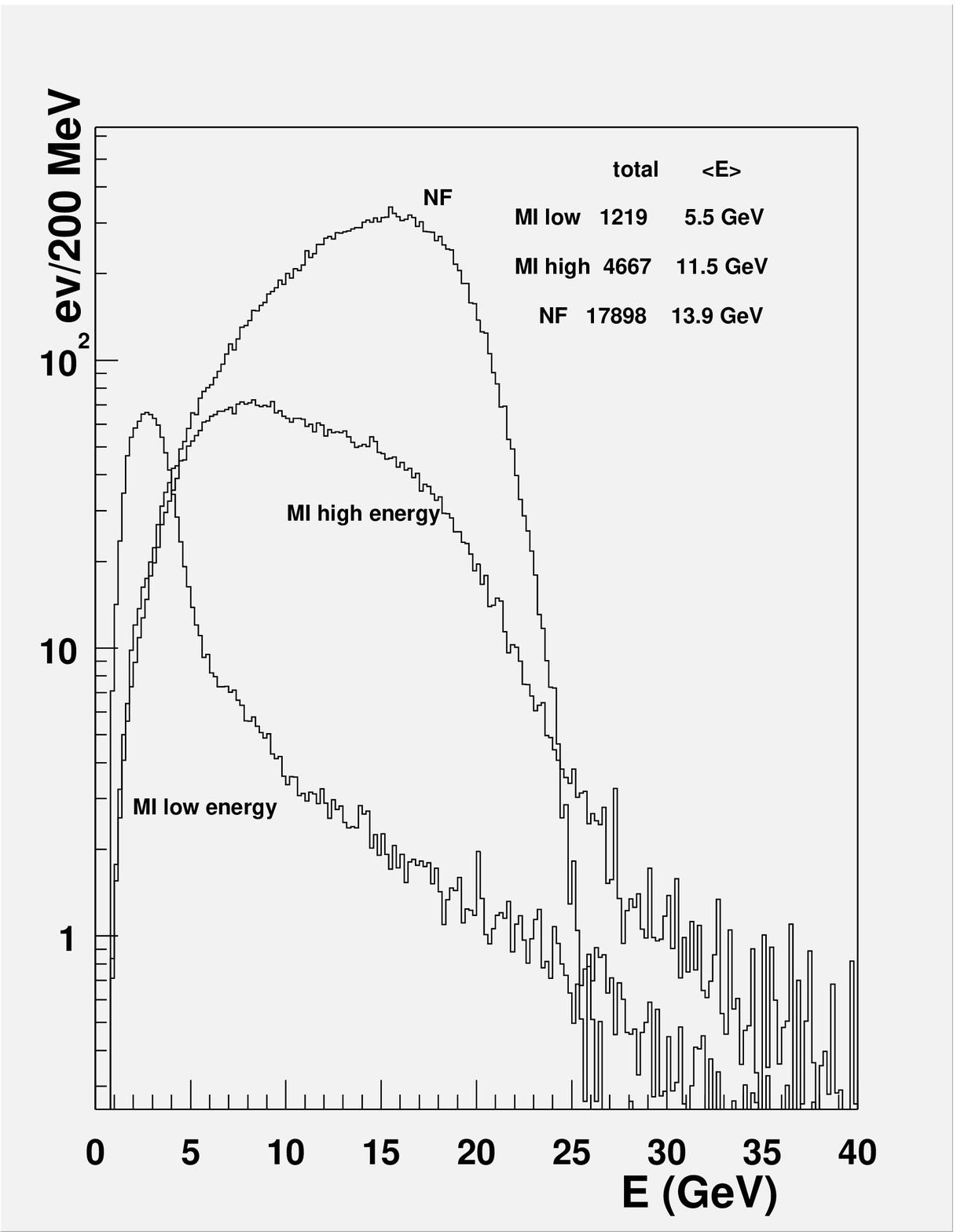}
\caption
{The oscillation-free "near" spectra of $\nu_\nu$-induced quasielastic 
events, $n(E)$, for the MINOS detector irradiated by the low-energy and 
high-energy beams from the Main Injector and by the beam from a neutrino 
factory storing 20-GeV negative muons. The assumed exposure in either beam 
is 10 kton--years.}
\label{spectra}
\end{figure}

\begin{figure}
\vspace{18 cm}
\includegraphics{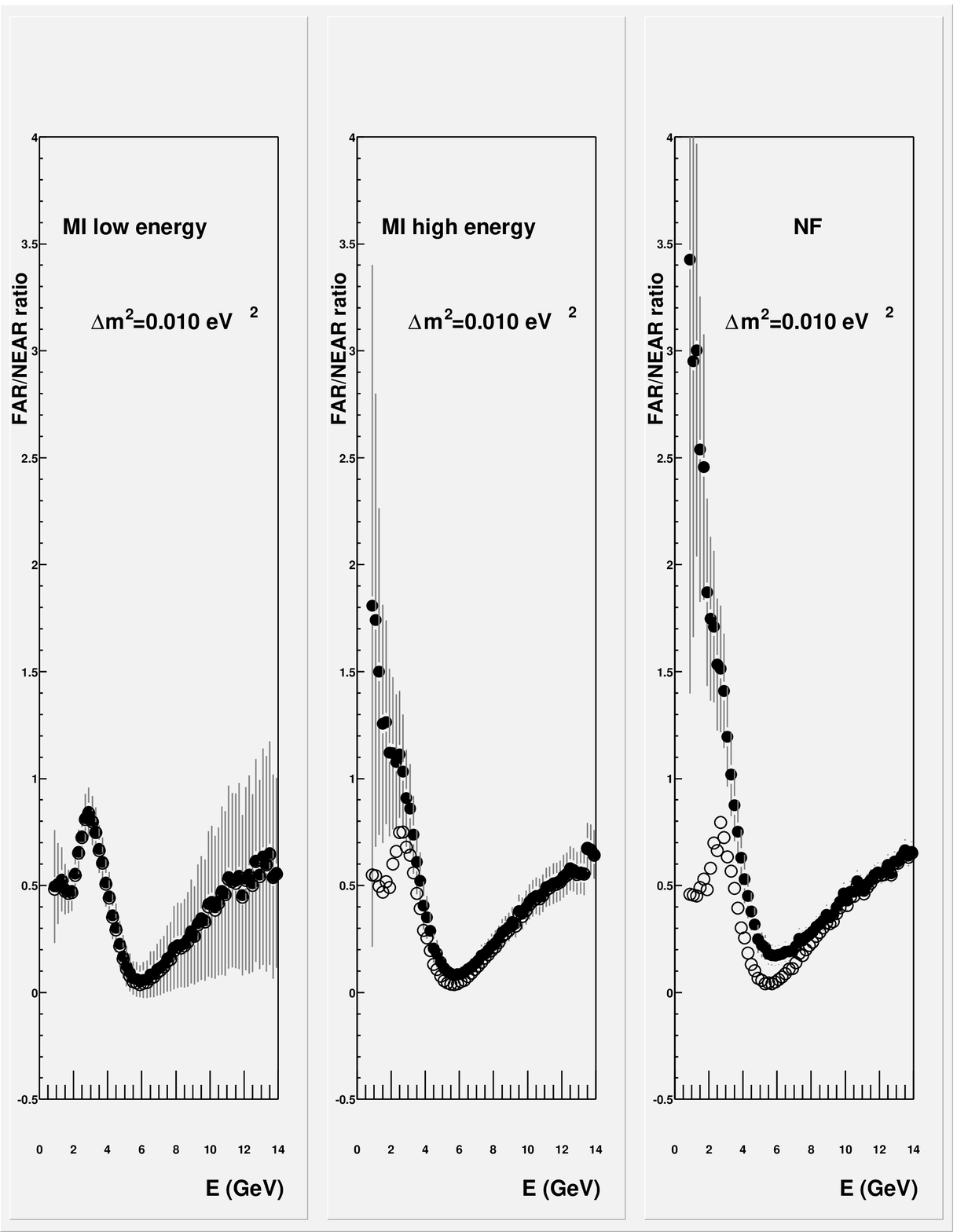}
\caption
{Assuming the transitions \omutau\ and \omuste\ (solid and open dots, 
respectively) driven 
by \dmsq\ = 10$^{-2}$ eV$^2$ and \mixing\ = 1, the far-to-near ratio
$R(E) = f(E) / n(E)$ for quasielastic events produced by the MI low-energy 
(a), MI high-energy (b), and NF (c) beams. The error bars on $R(E)$ for 
the transition \omutau\ are the statistical uncertainties corresponding to 
exposures of 10 kton--years in either beam.}
\label{ratios}
\end{figure}

\begin{figure}
\vspace{18 cm}
\includegraphics{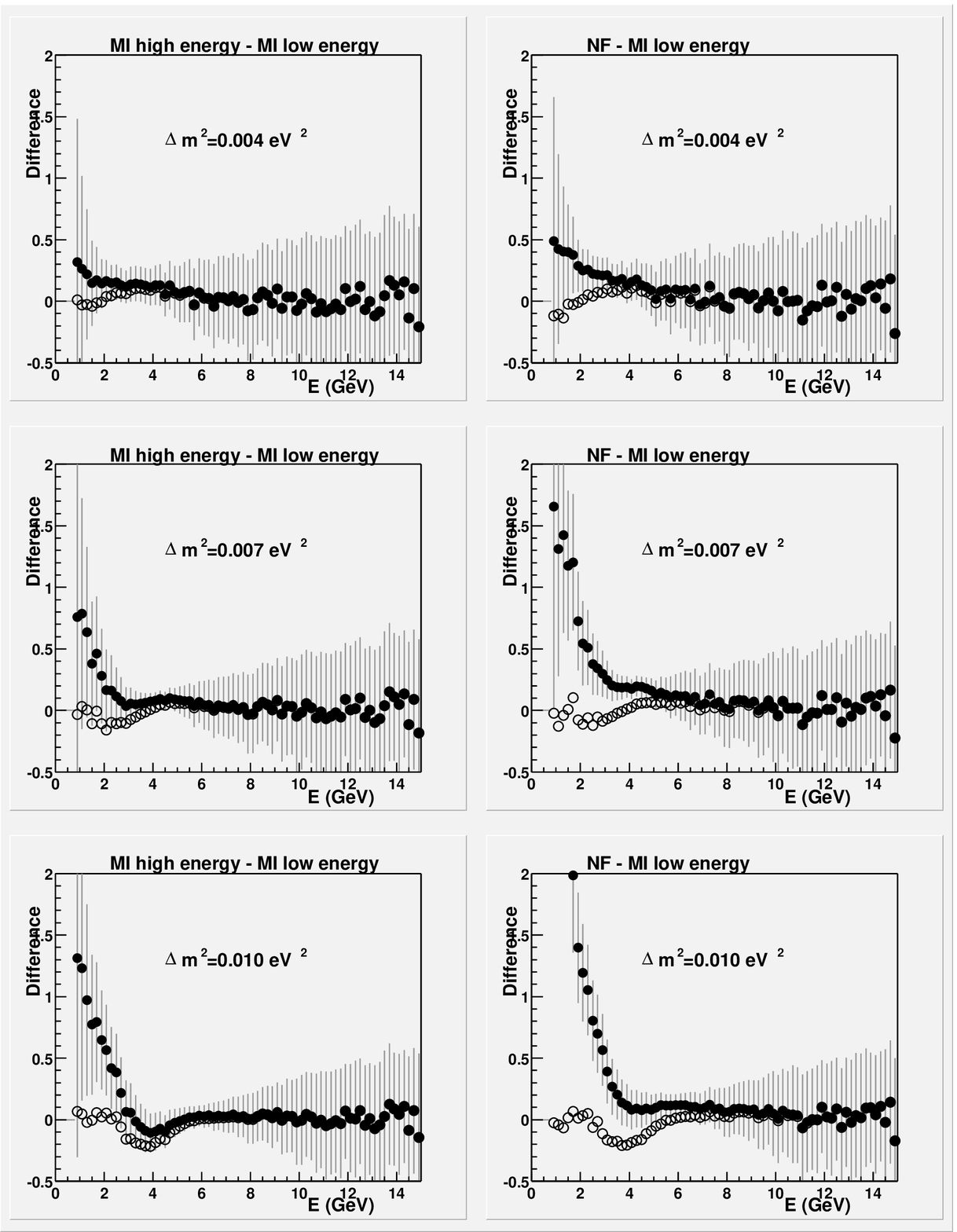}
\caption
{Assuming either \omutau\ or \omuste\ (solid and open dots, respectively)
with maximal mixing,
the difference $\Delta R(E)$ between the far-to-near ratios for: the MI
high-energy and low-energy beams (left-hand panels); the NF and MI low-energy 
beams (right-hand panels). The top, middle, and bottom panels are for 
\dmsq\ = 0.004, 0.007, and 0.010 eV$^2$, respectively. Depicted by error bars 
are the statistical errors that correspond to exposures of 10 kton--years in 
either beam.}
\label{difference}
\end{figure}

     Shown in Fig. \ref{spectra} are the oscillation-free near spectra of 
QE events, $n(E)$, for the three beams considered. In the absence of 
oscillations, equal exposures of 10 kton--years in the MI low-energy, MI 
high-energy, and NF beams will yield some 1200, 4600, and 17900
$\nu_\mu$-induced QE events, respectively\footnote{In all numerical 
estimates, we do not take into account the discussed upgrade of the proton 
driver at Fermilab \cite{driver} that may result in a substantial increase of 
neutrino flux from the Main Injector.}. Assuming either \omutau\ or 
\omuste\ driven by \dmsq\ = 0.01 eV$^2$ and maximal mixing of \mixing\ = 1, 
the far-to-near ratios $R(E)$ for the three beams are illustrated in 
Fig. \ref{ratios}\footnote{In the Figures, statistical fluctuations are 
suppressed for the data points themselves, but the error bars are for the 
statistics as indicated.}.
That the ratios $R(E)$ for the transitions \omutau\ and \omuste\ diverge 
towards low values of $E$ is evident for the MI high-energy and NF beams in 
which $\tau$ production is not suppressed. Always assigning the MI low-energy
beam as the reference beam and again considering \omutau\ and \omuste\ with 
maximal mixings, in Fig. \ref{difference} we plot the difference
\begin{displaymath}
\Delta R(E) = R_{\mathrm{hard}} - R_{\mathrm{soft}}
\end{displaymath}
for various values of \dmsq. Indeed, at visible energies below some 5 GeV
$\Delta R(E)$ deviates from zero for the transition \omutau, while staying
close to zero for \omuste. This deviation may be viewed as a signature of
$\nu_\tau$ appearance. The naive expectation for \omuste, $\Delta R(E) = 0$, 
is slightly violated by the smearing of neutrino energy.

\begin{figure}
\vspace{18 cm}
\includegraphics{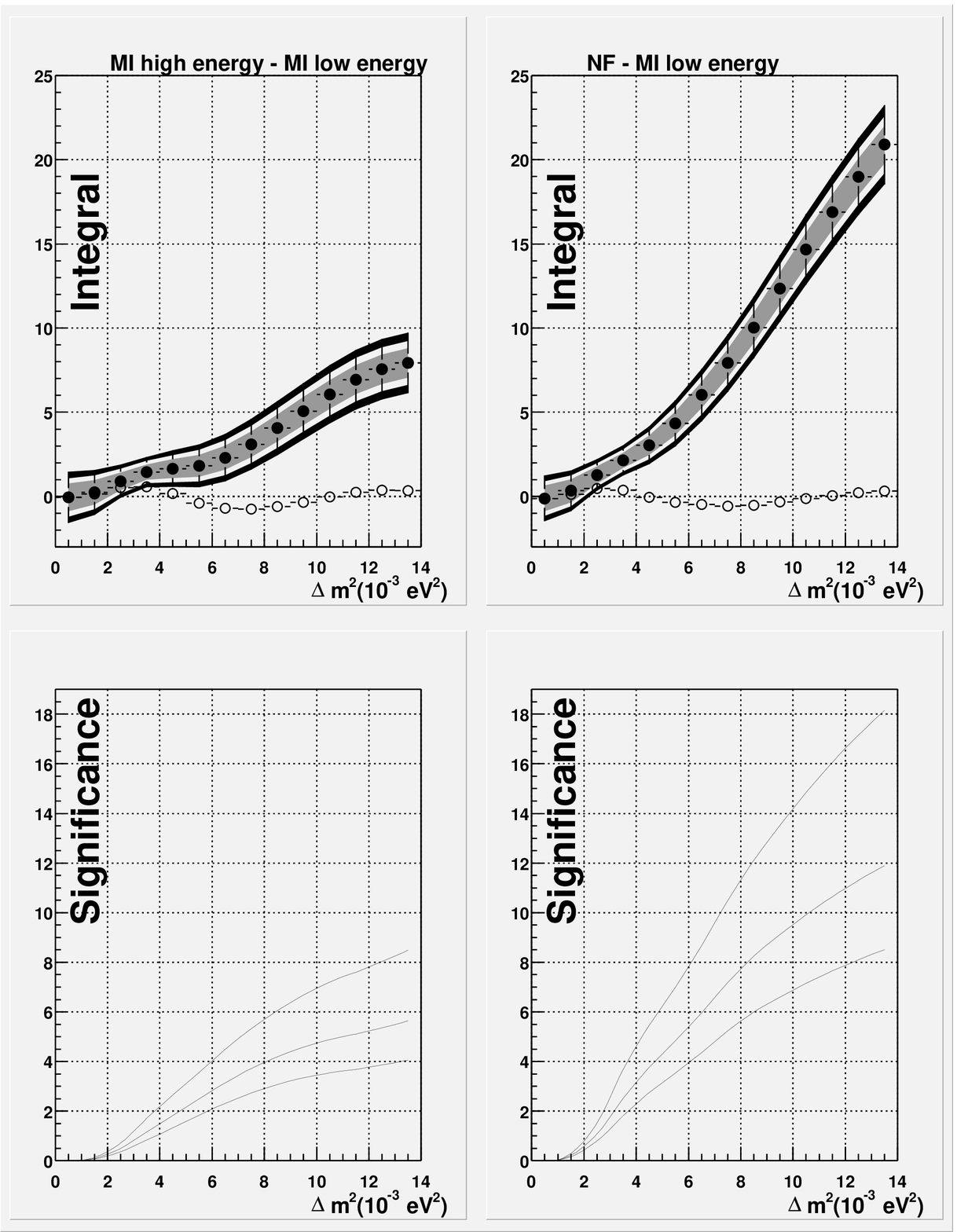}
\caption
{Assuming either \omutau\ or \omuste\ (solid and open dots, respectively)
with maximal mixing,
the integrated difference $S$ (see text) as a function of \dmsq\ for the 
MI high-energy (top left) and NF (top right) beams, with the MI low-energy
beam assigned as the reference beam in both cases. Shown by successive error 
corridors are the statistical errors on $S$(\dmsq) corresponding to exposures 
of 10, 20, and 50 kton--years in the $\tau$-producing beam and of 10 
kton-years in the reference beam. The bottom panels show the statistical 
significance of the $\tau$ signal for either case.}
\label{integral}
\end{figure}

     By the time the proposed test can be implemented, the actual value of
\dmsq\ will probably be estimated to some 10\% by analyzing the $\nu_\mu$
disappearance in the MI low-energy beam \cite{resolutions}. 
Given the value of \dmsq, a consistent approach would be to fit $\Delta R(E)$
to the predicted shape in order to estimate the mixing between the muon and 
tau neutrinos. A cruder measure of the effect is provided by the integral 
$S = \int \Delta R(E)dE$, which we estimate between $E = 1$ and 3 GeV.
Allowing for either \omutau\ and \omuste\ with maximal mixing, the respective
integrals $S$(\omutau) and $S$(\omuste) are plotted in Fig. \ref{integral} as
functions of \dmsq. 

     Since the selected reference beam produces many more low-energy events 
than the MI high-energy and NF beams, the statistical error on the integral 
$S$ is largely determined by the statistics accumulated with the harder beam.
Therefore, we fix the exposure in the reference beam at 10 kton--years and 
assume similar or bigger exposures in the MI high-energy and NF beams. The 
successive error corridors in Fig. \ref{integral} are the statistical 
uncertainties on $S$(\omutau) for the exposures of 10, 20, and 50 kton--years
in the latter beams. And finally, dividing the difference between 
$S$(\omutau) and $S$(\omuste) by the statistical error on $S$(\omutau), we 
estimate the statistical significance of the enhancement that is also 
depicted in Fig. \ref{integral}. We estimate that at a level of 3$\sigma$, 
the exposures of 10, 20, and 50 kton--years in the MI high-energy (NF) beam 
will allow to probe $\nu_\tau$ appearance down to the \dmsq\ values of some 
0.008 (0.005), 0.006 (0.004), and 0.005 (0.003) eV$^2$, respectively. 
Thus in the NuMI program with the existing Proton Booster \cite{numi},
the proposed test may be sensitive to \dmsq\ values in the Kamiokande-allowed
region \cite{kamioka}, but not below some $5\times 10^{-3}$ eV$^2$ as 
suggested by the more recent results of Super-Kamiokande \cite{superkamioka}.
Irradiating MINOS by a 20-GeV neutrino factory may allow to probe \dmsq\ 
values in the upper part of the region favored by Super-Kamiokande.

     To conclude, we have proposed a test of $\nu_\tau$ appearance that 
consists in comparing the far-to-near ratios of the spectra of quasielastic 
CC events generated by different beams of muon neutrinos, and therefore may 
be accessible to big calorimeters with muon spectrometry like MINOS. The 
test is limited by statistics rather than systematics, and its significance
crucially depends on the exposure in the harder beam in particular.

\clearpage

\end{document}